\documentclass[sigconf]{acmart}

\copyrightyear{2026}
\acmYear{2026}
\setcopyright{cc}
\setcctype{by}
\acmConference[CHIIR '26]{2026 ACM SIGIR Conference on Human Information Interaction and Retrieval}{March 22--26, 2026}{Seattle, WA, USA}
\acmBooktitle{2026 ACM SIGIR Conference on Human Information Interaction and Retrieval (CHIIR '26), March 22--26, 2026, Seattle, WA, USA}
\acmPrice{}
\acmDOI{10.1145/3786304.3787876}
\acmISBN{979-8-4007-2414-5/2026/03}
\begin{document}


\title{What LLMs Miss in Recommendations: Bridging the Gap with Retrieval-Augmented Collaborative Signals}
\title{Structured Prompts for LLMs: Towards Better Personalized Suggestions}
\title{What LLMs Miss in Recommendations: Bridging the Gap with Retrieval-Augmented Collaborative Signals}
\title{Empowering LLMs with User-Item Interactions: A Retrieval-Augmented Approach}
\title{Do LLMs Understand Collaborative Signals? Diagnosis and Repair}

\title{What LLMs Miss in Recommender Systems?}
\title{Recommender Reasoning in LLMs: What's Missing?}

\title{Can LLMs Learn to Recommend? Diagnosing Collaborative Failures}
\title{Do LLMs Understand Collaborative Signals? Diagnosis and Repair}
\author{Shahrooz Pouryousef}
\affiliation{%
  \institution{UMass Amherst}
  \streetaddress{1 Th{\o}rv{\"a}ld Circle}
  \city{Amherst, MA}
  \country{USA}}
\email{shahrooz@cs.umass.edu}

\author{Ali Montazer}
\affiliation{%
  \institution{Google}
  \streetaddress{1 Th{\o}rv{\"a}ld Circle}
  \city{Mountain View, CA}
  \country{USA}}
\email{alimontazer@google.com}







\providecommand{\myparab}[1]{\smallskip\noindent\textbf{#1} }
\begin{abstract}
Collaborative information from user-item interactions is a fundamental source of signal in successful recommender systems. Recently, researchers have attempted to incorporate this knowledge into large language model-based recommender approaches (LLMRec) to enhance their performance. However, there has been little fundamental analysis of whether LLMs can effectively reason over collaborative information. In this paper, we analyze the ability of LLMs to reason about collaborative information in recommendation tasks, comparing their performance to traditional matrix factorization (MF) models. We propose a simple and effective method to improve LLMs' reasoning capabilities using retrieval-augmented generation (RAG) over the user-item interaction matrix with four different prompting strategies.
Our results show that the LLM outperforms the MF model whenever we provide relevant information in a clear and easy-to-follow format, and prompt the LLM to reason based on it. We observe that with this strategy, in almost all cases, the more information we provide, the better the LLM performs. 
\end{abstract}





\maketitle

\section{Introduction}
\vspace{-0.05in}
Large Language Models (LLMs) \cite{achiam2023gpt, touvron2023llama} have demonstrated impressive capabilities in a wide range of tasks, including reasoning over textual input \cite{huang2022towards, plaat2024reasoning}, answering complex questions \cite{singhal2025toward, jiang2021can}, generating fluent text \cite{gao2023enabling}, and encoding world knowledge \cite{singhal2023large}. As a result, LLMs have been increasingly adopted in diverse research domains such as data collection \cite{wei2024llmrec}, summarization \cite{basyal2023text, liu2023learning}, translation \cite{zhu2023multilingual, kocmi2024findings, raunak2023leveraging}, and visualization \cite{li2024visualization, wu2024automated}.

More recently, researchers have begun exploring the potential of LLMs in recommender systems, where the goal is to deliver personalized item suggestions that align with a user's preferences—such as recommending movies, products, or articles \cite{bao2023tallrec, dai2023uncovering, sanner2023large, yue2023llamarec}. This emerging line of work, often referred to as LLMs as recommenders (LLMRec), presents an exciting new research direction that leverages the capabilities of LLMs to tackle core challenges in recommender systems \cite{zhao2024recommender, bao2023tallrec, zhang2023chatgpt}. Within this domain, LLMs have primarily been applied to two key subtasks: item understanding and user modeling. For example, in content-based filtering, LLMs can generate rich item representations from textual descriptions and infer user preferences from their interaction history \cite{lin2025can, wu2024survey, liao2024llara}.

LLMs have been utilized in recommender systems (RSs) under two main paradigms. LLMs as Recommenders (LLMs-as-RSs) refers to approaches where LLMs are directly prompted or fine-tuned to function as recommenders. In contrast, LLM-enhanced RSs leverage the knowledge stored in LLM parameters to enhance traditional recommender models—for example, by generating text-based item/user representations or embeddings.

One limitation of LLMRec methods is their insufficient modeling of collaborative information embedded in user-item co-occurrence patterns. To address this, \citet{sun2024large} proposed an approach that distills the world knowledge and reasoning capabilities of LLMs into a collaborative filtering recommender system. Their method adopts an in-context, chain-of-thought prompting strategy, focusing on the LLM-enhanced RS paradigm.

However, the ability of LLMs to reason over collaborative filtering information remains limited. Recent work (e.g., \cite{kim2024large}) attempts to bridge this gap by aligning user–item embeddings with the LLM’s representation space, but such methods still fall short—LLMs cannot directly access or reason over the full set of similar users’ interactions due to input and context constraints.

In this paper, we take a step toward addressing this gap by analyzing the reasoning capabilities of LLMs over collaborative signals. Our first research question is: \textit{Can LLMs capture and reason over collaborative patterns when provided with necessary information, such as similar users or items?} To examine this, we compare LLMs with matrix factorization—a classical and widely used method for modeling collaborative information in recommender systems. This comparison enables us to assess whether LLMs can reason over collaborative signals at a level comparable to this simple yet strong baseline.

Our second research question examines \textit{how to enhance LLMs’ ability to reason over collaborative data.} To address the challenges discussed above, we explore a retrieval-augmented generation (RAG) approach that provides the LLM with relevant user–item interaction information at inference time, aiming to bridge the gap between LLMs and traditional collaborative filtering methods. With this strategy, we reduce noise in the LLM’s input and help it reason with greater precision.

We examine four different prompting strategies and show that providing relevant information in a clear format and prompting the LLM to reason over it significantly improves performance, especially for cold-user data. We also show that, given relevant information, the LLM can effectively continue making recommendations as more context becomes available—for example, from additional similar users or more items associated with them.

\section{Preliminaries}
\vspace{-0.05in}
\begin{figure}
\centering
    \includegraphics[width=0.48\textwidth]{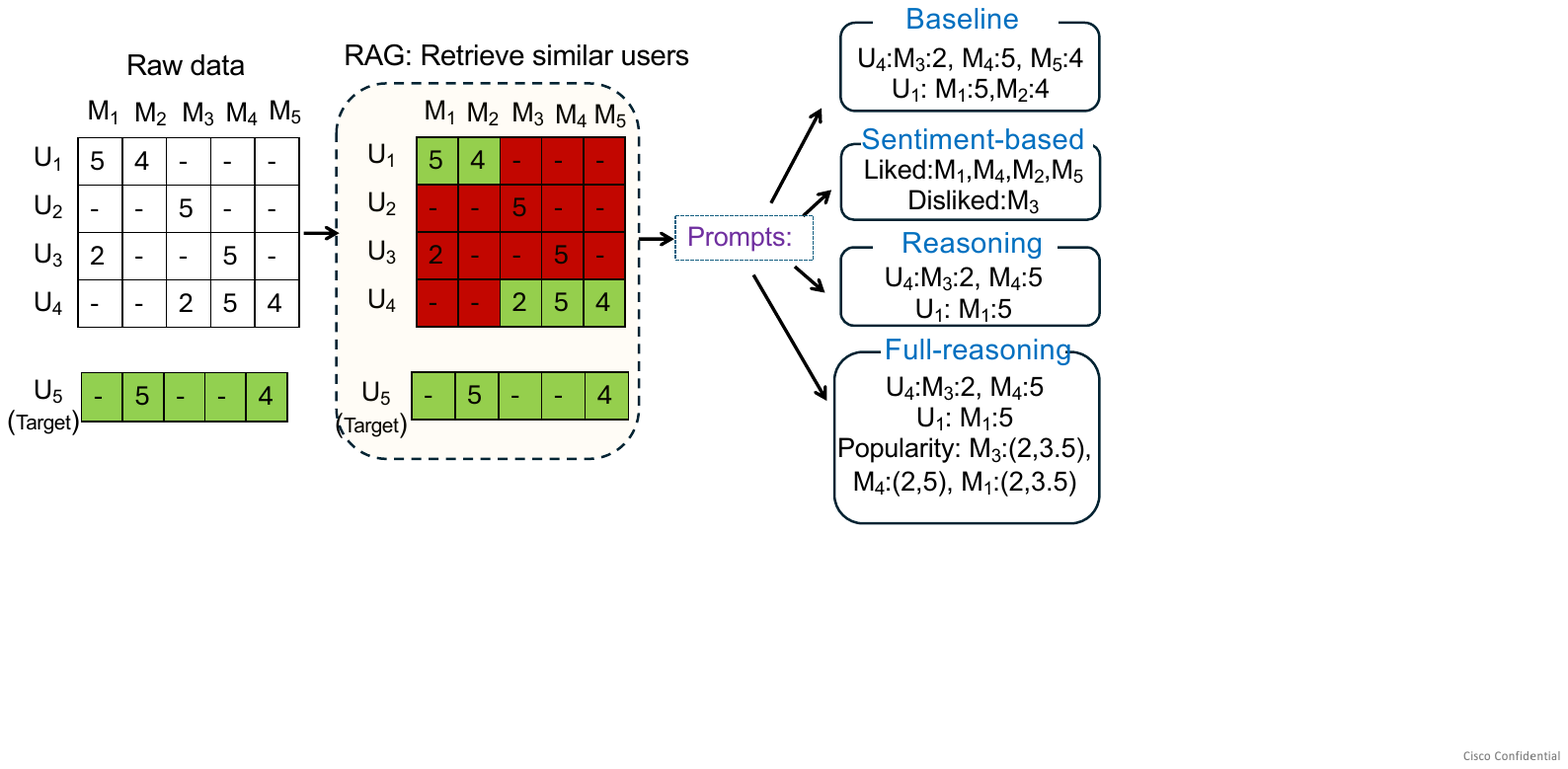}
    \vspace{-0.3in}
  \caption{Comparison of four prompt-generation strategies for movie recommendation based on retrieved user–movie similarities. Each method varies in how it incorporates similar users’ ratings, handles previously seen movies, and structures the prompt for downstream recommendation.}
  \label{fig:example}
\end{figure}
In this study, we explore the reasoning ability of LLMs on collaborative filtering information. Our investigation is centered around a matrix of interactions between users and items, i.e., \( M \). Let \( \mathcal{U} = \{u_1, \dots, u_n\} \) denote the set of all \( n \) users and \( \mathcal{I} = \{v_1, \dots, v_m\} \) denote the set of all \( m \) items (movies). The matrix \( M \in \mathbb{R}^{n \times m} \) represents the rating behavior of users over items, where each row corresponds to a user in \( \mathcal{U} \) and each column to an item in \( \mathcal{I} \) as shown in the left side matrix in figure \ref{fig:example}. The entry \( M_{ij} \) indicates the rating that user \( u_i \in \mathcal{U} \) has given to item \( v_j \in \mathcal{I} \), with values typically ranging from \( 1 \) (disliked) to \( 5 \) (highly liked), or \( 0 \) if the user has not rated the item.

To enable LLMs to understand collaborative signals in the interaction matrix, it is necessary to provide them with structured information derived from the matrix. However, including the entire matrix in the input is infeasible, especially in large-scale datasets with many users and items. Moreover, presenting all such information in a prompt can overwhelm the LLM, making it difficult to interpret the collaborative structure and respond effectively to recommendation queries.

Recently, retrieval-augmented generation (RAG) has demonstrated impressive results in enhancing LLM performance across a variety of tasks \cite{fan2024survey, chen2024benchmarking, zhang2024r}. RAG approaches integrate external retrieval mechanisms to supplement LLMs with relevant and targeted information, thereby improving the accuracy, relevance, and contextual understanding of generated responses.

In the following section, we present our method for applying RAG to enhance the effectiveness and efficiency of LLMs in leveraging collaborative filtering information.

\section{Methodology}
\vspace{-0.05in}
Given a target user $u_t$, our goal is to extract relevant signals from the interaction matrix $M$ to enable an LLM to predict and recommend movies that the user is likely to like. To identify useful information, we first retrieve users most similar to the target user. These similar users are assumed to carry the most informative signals for inferring the preferences of $u_t$, as their past ratings can help guide the recommendation of unseen items.

To identify users most similar to the target user \( u_t \), we compute pairwise user similarities using cosine similarity over their rating vectors: 
: 
\[
\text{sim}(u_t, u) = \frac{M_{u_t} \cdot M_u}{\|M_{u_t}\| \cdot \|M_u\|}
\]
where \( M_{u_t} \) and \( M_u \) denote the rating vectors of users \( u_t \) and \( u \), respectively. Each vector is constructed from the corresponding row in the interaction matrix \( M \), with unobserved ratings treated as zeros. We define the set of top-\(k\) similar users as:

\[
\mathcal{N}_k(u_t) = \text{Top-}k \left( \{ \text{sim}(u_t, u) \mid u \in \mathcal{U} \setminus \{u_t\} \} \right)
\]
where \( \mathcal{U} \) is the set of all users, and \( \mathcal{N}_k(u_t) \subset \mathcal{U} \setminus \{u_t\} \) denotes the \( k \) most similar users to \( u_t \). 
For each \( u \in \mathcal{N}_k(u_t) \), we define their set of ratings as:

\[
R_u = \{ (j, M_{u_j}) \mid j \in \mathcal{I},\ M_{u_j} \ne 0 \}
\]

We use the information from these similar users and their ratings to construct the input for the LLM in our prompt-generation framework. Specifically, the sets \( R_u \) for all \( u \in \mathcal{N}_k(u_t) \) provide the relevant collaborative signals that are encoded into the prompt. Let \( \mathcal{R}_{\mathcal{N}_k(u_t)} = \bigcup_{u \in \mathcal{N}_k(u_t)} R_u \) denote the union of all such rating sets. This set \( \mathcal{R}_{\mathcal{N}_k(u_t)} \subset \mathcal{I} \times \mathbb{R} \) represents the pool of item–rating pairs from users most similar to the target user \( u_t \), and serves as a compact, informative context for the LLM.

In the following subsections, we present four distinct strategies for incorporating \( \mathcal{R}_{\mathcal{N}_k(u_t)} \) into the prompt. Each strategy varies in how it structures this information, filters relevant ratings, and balances the trade-off between informativeness and prompt length.

\subsection{Prompt Generation Strategies}
\vspace{-0.05in}
We explore four distinct strategies for constructing our prompts. Each approach presents user rating data differently to investigate its impact on model performance and prompt efficiency. We explain this approach and provide one example of each strategy in figure \ref{fig:example}.

\subsubsection{\textbf{Unfiltered Full Ratings Prompt (Baseline)}}

This baseline incorporates the complete rating histories of the top-$k$ most similar users without applying any filtering or prioritization. Each user's ratings are included in their original, unprocessed form. Although the fraction parameter ($f$) still determines the proportion of similar users’ data included in the prompt, the selection does not favor unseen items or those with higher ratings. By presenting all available preference data, this approach aims to provide a broad and potentially richer context to the LLM.

Importantly, this approach includes all sampled ratings from similar users, even for items that the target user has already rated. This design choice allows the LLM to observe both the target user's and similar users' opinions on the same items, potentially helping it better infer the target user's preferences. At the same time, we explicitly instruct the LLM not to select movies that the target user has already rated. We refer to this strategy as our \textit{baseline}.

\subsubsection{\textbf{Sentiment-Based Prompt}}

In this strategy, ratings from similar users are grouped into three sentiment categories: \textit{Liked}, \textit{Neutral}, and \textit{Disliked}. These categories are determined using rating thresholds: ratings $\geq 4$ are considered \textit{Liked}, a rating of $3$ is treated as \textit{Neutral}, and ratings $\leq 2$ fall under \textit{Disliked}. Additionally, we exclude any movies that the target user has already rated to ensure the model focuses solely on unseen content.






\begin{figure*}
\centering
    \includegraphics[width=17cm]{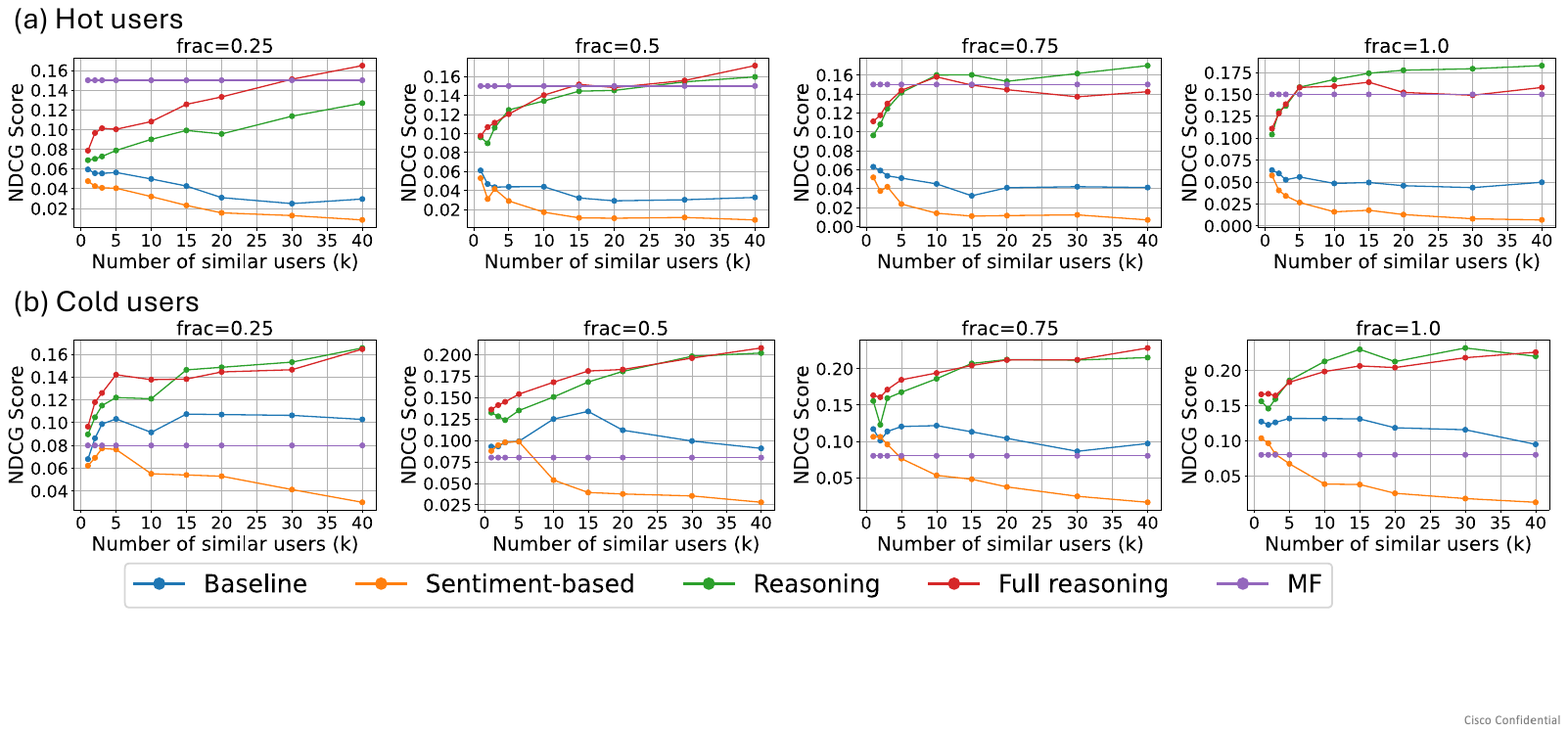}
    \vspace{-0.2in}
  \caption{NDCG score for different prompt generation strategies and MF as a function of $k$ and $f$. }\label{fig:NDCG_as_prompt_generation_methods}
\end{figure*}

\subsubsection{\textbf{Reasoning-Based Prompt}}

The interaction matrix in collaborative filtering captures valuable relational signals that can be interpreted as chains of reasoning, leading to meaningful recommendations. For example, if user A and user B both rated a particular item as 2, we can infer that their preferences are similar. If user B then gives a rating of 5 to another item, it is reasonable to predict that user A may also enjoy that item.

While this example involves just two users and two items, the underlying idea extends to more complex scenarios involving many users and items. This kind of transitive inference demonstrates how collaborative signals can be composed to uncover deeper and more nuanced patterns of user preferences.

Inspired by the success of chain-of-thought prompting for LLMs \cite{wei2022chain}, we design this strategy to encourage structured reasoning over such collaborative chains. Specifically, we build on the baseline by introducing two key modifications. First, we exclude any movies from similar users’ ratings that have already been rated by the target user, ensuring that all recommended items are truly unseen.
Second, we explicitly frame the prompt as a reasoning task by appending a directive such as: \textit{“Reason based on the patterns above: which 10 movies should user A watch next that they haven't seen?”} This formulation encourages the LLM to perform inference based on user-item interaction patterns, rather than relying purely on shallow matching. 

By structuring the task in this way, we implicitly apply a “step-by-step” reasoning strategy, which has been shown to improve the performance of LLMs across a range of tasks \cite{wei2022chain}.

\vspace{-0.1cm}
\subsubsection{\textbf{Full Reasoning}}

In the \textit{Reasoning} approach, the LLM is exposed only to the preferences of the top-$k$ most similar users and therefore lacks access to global information—such as how popular a movie is across the entire user base. To address this limitation, we introduce an enhanced variant that augments the prompt with additional popularity-based statistics.

For each movie \( v_j \in \mathcal{I} \), we compute two global metrics derived from the interaction matrix \( M \in \mathbb{R}^{n \times m} \):

\begin{itemize}
    \item \textbf{Rating Count:} The number of users who have rated movie \( v_j \):
    \[
    \text{Count}(v_j) = \left| \left\{ u_i \in \mathcal{U} \mid M_{ij} \ne 0 \right\} \right|
    \]
    
    \item \textbf{Average Rating:} The mean rating received by movie \( v_j \):
    \[
    \text{AvgRating}(v_j) = \frac{\sum_{u_i \in \mathcal{U}} M_{ij} \cdot \mathbf{1}[M_{ij} \ne 0]}{\text{Count}(v_j)}
    \]
\end{itemize}

These statistics are appended to each corresponding movie entry in the \textit{Reasoning} prompt. By doing so, the LLM is equipped to weigh both local signals (from similar users) and global signals (reflecting overall popularity and quality), potentially improving the relevance and diversity of its recommendations.

\section{Evaluation}
\vspace{-0.05in}

\subsection{Dataset}
\vspace{-0.05in}
We evaluate our methods and baselines using the widely studied MovieLens 100K dataset \cite{harper2015movielens}, which contains 100,000 ratings provided by 943 users on 1,682 movies. As part of preprocessing, we remap both user and movie IDs to contiguous integers starting from zero to ensure compatibility with our model input formats.

To simulate realistic recommendation scenarios, we sort each user's rating history chronologically and randomly mask 20\% of their ratings to represent unobserved preferences. These masked ratings are withheld during training and later used for evaluation. The remaining 80\% of the ratings are treated as known history and are used to compute user-user similarities and construct prompts for the language model.

\begin{figure*}
\centering
    \includegraphics[width=17cm]{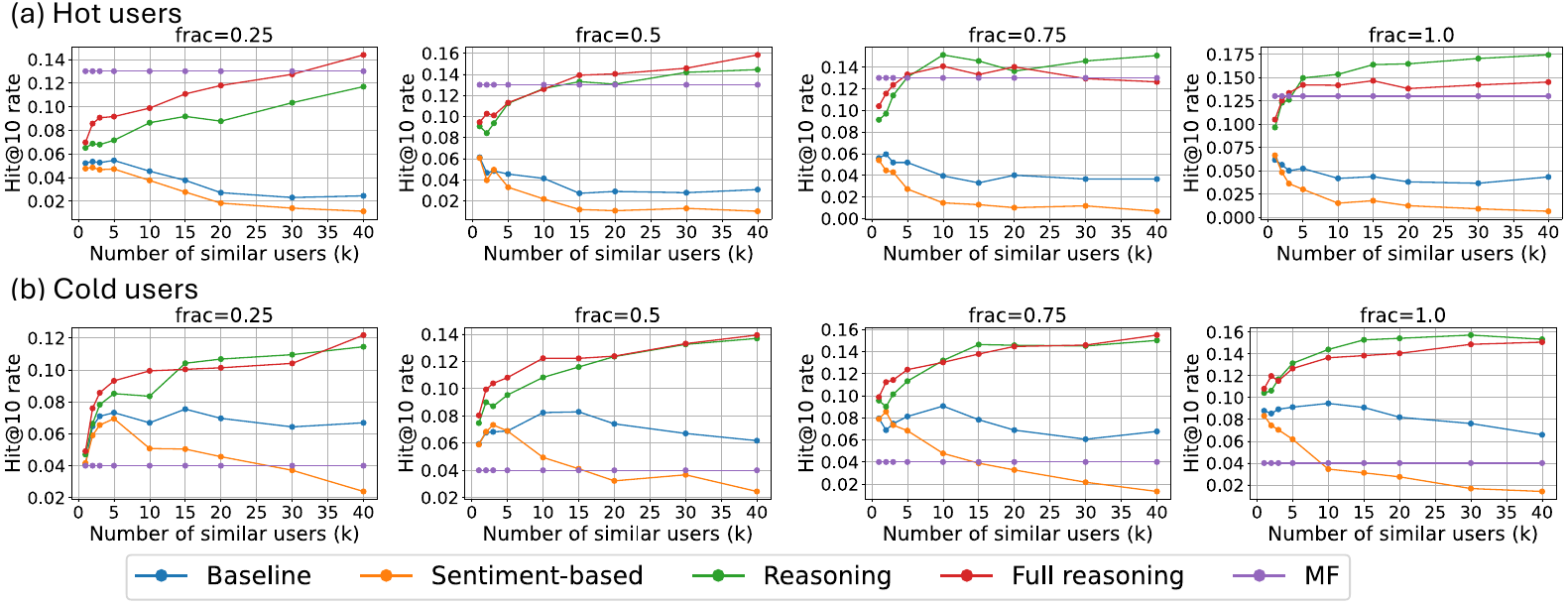}
    \vspace{-0.2in}
  \caption{Hit@10 score for different prompt generation strategies and MF as a function of $k$ and $f$.}\label{fig:hit@10_as_prompt_generation_methods}
\end{figure*}

To ensure fair evaluation across varying user profiles, we partition users into two groups based on their activity level. Specifically, users with a number of ratings above the dataset’s median are labeled as \textit{hot users}, while those below or equal to the median are considered \textit{cold users}. We randomly sample 350 users from each group to construct balanced evaluation sets.
 
\vspace{-0.05in}
\subsection{Experimental and parameter settings}
\vspace{-0.05in}
To simulate varying levels of auxiliary knowledge, we control the fraction $f$ (25\% to 100\%) of each similar user's ratings included in the prompt, allowing us to examine how contextual information impacts recommendation quality. We also inform the LLM of the valid movie ID range to help it interpret the data consistently. Each prompt ends by asking the model to recommend 10 unseen movies for the target user. We query the GPT-4.1-mini model via OpenAI’s ChatCompletion API and extract recommended movie IDs from its response. These are evaluated using Normalized Discounted Cumulative Gain (NDCG) \cite{jarvelin2002cumulated}, which rewards correctly ranked relevant items, and Hit@10, which measures how often ground-truth masked movies appear in the top 10 results.

\vspace{-0.1in}
\subsection{Results and Discussion}
\vspace{-0.05in}
Figures~\ref{fig:NDCG_as_prompt_generation_methods} and~\ref{fig:hit@10_as_prompt_generation_methods} show average NDCG and Hit@10 scores across prompt generation strategies, varying the number of similar users 
$k$ and fraction $f$ of their ratings included. Reasoning-based prompts consistently improve with larger 
$k$, indicating LLMs benefit from structured, interpretable collaborative signals. In contrast, simpler strategies (e.g., sentiment-based or original) decline as 
$k$ increases, likely due to unstructured raw data overwhelming the model. These findings underscore the need for structured prompt design, potentially leveraging RAG or knowledge graphs.

As the fraction of similar user data in the prompt increases, NDCG scores improve for reasoning strategies, confirming that more context helps LLM predictions. Notably, LLMs perform better on cold users, while Matrix Factorization (MF) favors hot users due to its reliance on rich interaction histories for learning effective embeddings.

For LLMs, one possible explanation for stronger performance on cold users is reduced confusion due to fewer input ratings. Additionally, cold users typically have fewer masked (unseen) movies—sometimes only 2 or 3—since we mask 20\% of each user's ratings. When the LLM is asked to recommend 10 movies, the probability of hitting those few masked movies is relatively high. In contrast, hot users often have 100+ ratings, so 20\% masking can result in 20–30 unseen movies. Consequently, the LLM has a lower chance of selecting the correct ones, even when reasoning is good, due to the larger candidate set.

\begin{figure}[t]
\centering
    \includegraphics[width=7.3cm]{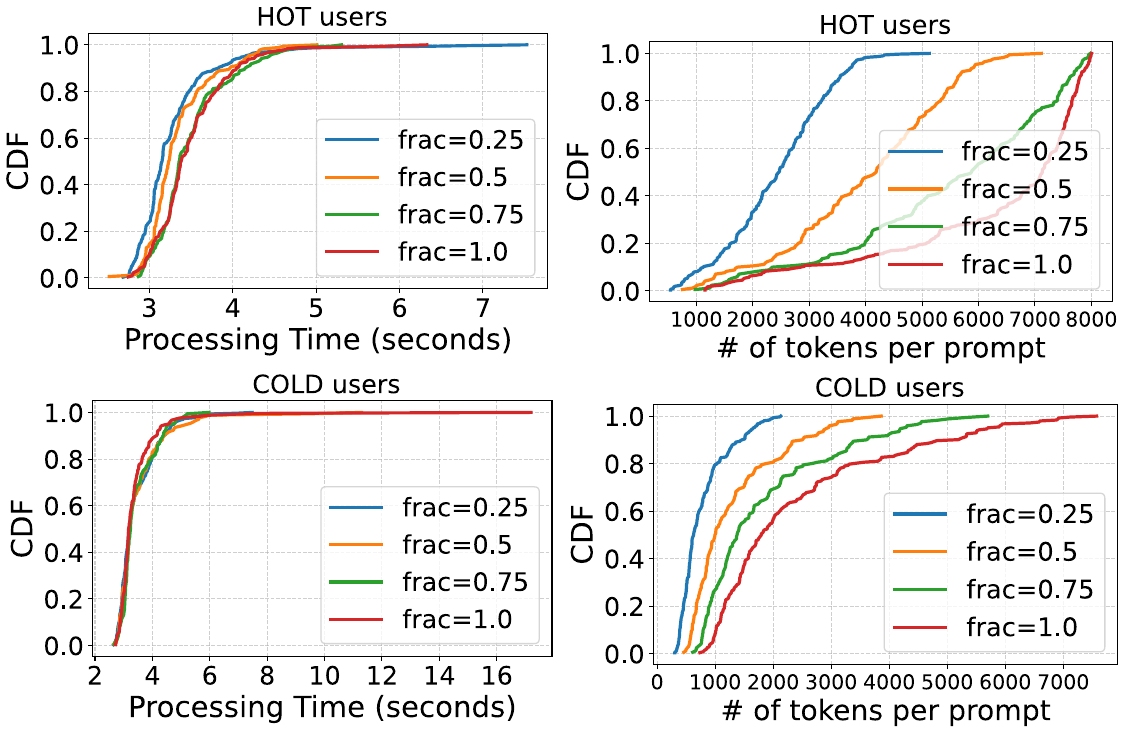}
    \vspace{-0.15in}
  \caption{Processing time of prompts and the number of tokens in each prompt for hot and cold users.}\label{fig:CDF_time_tokens}
\end{figure}

Figure~\ref{fig:CDF_time_tokens} shows the CDF of prompt processing times and prompt sizes (number of tokens) for cold and hot users using the Full reasoning strategy, with $k=10$ and varying fractions of each similar user’s ratings. Processing time includes prompt transmission, GPT computation, and response retrieval. For cold users, CDF curves across fractions ($0.25-1.0$) are nearly identical, as they have few ratings and adding more from similar users does not significantly increase prompt length or latency. While increasing the fraction does slightly increase token count for cold users, their prompts remain shorter than those of hot users. In contrast, hot users show clear increases in processing time with higher fractions-larger histories lead to longer prompts and higher latency, evident from the rightward shift in the CDF curves. These findings highlight that prompt length strongly impacts latency for hot users and emphasize the need for token-efficient prompt strategies in scalable LLM-based recommendation systems.

\vspace{-0.05in}
\section{Conclusion}
\vspace{-0.05in}
In the LLMs-as-recommender-systems (LLMs-as-RCs) paradigm, a key challenge is enabling LLMs to effectively incorporate collaborative information. In this paper, we began by analyzing the performance of LLMs in capturing collaborative signals for movie recommendation. We showed that a naive approach—embedding all user information directly into the prompt—makes it difficult for LLMs to interpret these signals, often performing worse than simple baselines such as matrix factorization. To address this, we proposed a RAG based approach with improved prompting strategies. Our results demonstrate that presenting collaborative signals in a compact format and prompting the LLM to reason over them improves recommendation performance compared to traditional baselines. Moreover, our method is both token-efficient and effective, achieving better results on standard evaluation metrics while minimizing prompt length.


\bibliographystyle{ACM-Reference-Format}
\bibliography{references.bib}

\end{document}